\documentclass[twocolumn,secnumarabic,amssymb, amsmath, nofootinbib,tightenlines,
nobibnotes, aps, prl,epsfig]{revtex4}
\usepackage{graphicx}
\usepackage{dcolumn}
\usepackage{bm}
\begin{document}
\preprint{APS/123-QED}
\title{Non-linear corrections to the longitudinal structure function $F_{L}$ from the parametrization of $F_{2}$: Laplace transform approach }

\author{G.R.Boroun}%
 \email{grboroun@gmail.com; boroun@razi.ac.ir }
\affiliation{Department of Physics, Razi University, Kermanshah
67149, Iran}
\date{\today}
\begin{abstract}
The non-linear corrections (NLC) to the longitudinal structure
function in a limited approach is derived at low values of the
Bjorken variable $x$ by using the Laplace transforms technique.
The non-linear behavior of the longitudinal structure function is
determined with respect to the Gribov-Levin-Ryskin Mueller-Qiu
(GLR-MQ) and Altarelli-Martinelli (AM) equations. These results
show that the non-linear longitudinal structure function can be
determined directly in terms of  the parametrization of
$F_{2}(x,Q^{2})$ and the derivative of the proton structure
function with respect to $\ln{Q^{2}}$. These corrections improve
the behavior of the longitudinal structure function at low
values of $Q^{2}$ in comparison with other parametrization methods.\\
\end{abstract}
 \pacs{***}
\maketitle
The recent articles [1,2] revive the longitudinal structure
function $F_{L}(x,Q^{2})$ due to the parametrization of
$F_{2}(x,Q^{2})$ [3] and its derivative at low values of $x$. In
these papers (i.e., [1] and [2]) authors show that the
longitudinal structure functions have been obtained in the Mellin
and Laplace transform techniques from the parametrization of
$F_{2}(x,Q^{2})$ respectively. These results at leading-order
approximation lead to the unique results in a wide range of
$Q^{2}$ values. Theoretical analysis of these results at low $x$,
in context of fulfilment of the Froissart boundary, is of a great
importance in ultra-high energy processes in the future
 electron-proton colliders. Parametrization of the proton
 structure function has suggested by authors in Ref.[3], which
 describe fairly well the experimental data [4] at low $x$  in a wide range of the
momentum transfer $Q^{2}$, $x{\leq}0.1$ and
$0.15~\mathrm{GeV}^{2}<Q^{2}<3000~\mathrm{GeV}^{2}$. Ref.[2] shows
that the description of the data can be improved by allowing for
phenomenological rescaling variable corrections to the Bjorken
scaling variable. They are due to the charm quark mass (for
$n_{f}=4$) by the following form
$\chi=x(1+\frac{4m_{c}^{2}}{Q^{2}})$, with
$m_{c}=1.29^{+0.077}_{-0.053} \mathrm{GeV}$ [5]. Altogether the
results in Refs.[1] and [2] contain good agreement with the
experimental data at large $Q^{2}$ for the longitudinal structure
functions, but at low $Q^{2}$ these results can be improved when
take into account the non-linear corrections.\\
It is known that in the low $x$ and low $Q^{2}$ regions, the
non-linear corrections (or gluon recombination effects) are not
negligible and reduce the growth of the gluon distribution
function [6-9]. The non-linear corrections of gluon recombination
to the parton distributions have been calculated by Gribov-Levin-
Ryskin (GLR) and Mueller-Qiu (MQ) in [10] based on the
Abramovsky-Gribov-Kancheli (AGK) cutting rules in the double
leading logarithmic approximation (DLLA). The study of the GLR-MQ
equation may provide important insight into the non-linear
corrections of gluon recombination due to the high gluon density
at sufficiently low $x$. Indeed the number of partons in a phase
space cell $({\Delta}\ln(1/x){\Delta}\ln{Q^{2}})$ increases
through gluon splitting and decreases through gluon recombination.
That is, all possible $g+g{\rightarrow}g$ ladder recombinations
are resummed to leading order of the parameter
$\alpha_{s}\ln(1/x)\ln(Q^{2}/Q^{2}_{0})$. It leads to saturation
of the gluon density at low $Q^{2}$ with decreasing $x$. Thus the
non-linear corrections emerged from the recombination of two gluon
ladders, modify the evolution equations of singlet quark
distribution. Indeed an extra  non-linear term added to the linear
DGLAP evolution equation by the following form
\begin{eqnarray}
\frac{\partial{F_{2}(x,Q^{2})}}{\partial{\ln}Q^{2}}&=&\frac{\partial{F_{2}(x,Q^{2})}}{\partial{\ln}Q^{2}}|_{DGLAP}\\
&&-<e^{2}>\zeta[xg(x,Q^{2})]^{2}+HT.\nonumber
\end{eqnarray}
Here, the representation for the gluon distribution
$G(x,Q^{2})=xg(x,Q^{2})$ is used and  $<e^{2}>$ is the average of
the charge $e^{2}$ for the active quark flavors,
$<e^{2}>=n_{f}^{-1}\sum_{i=1}^{n_{f}}e_{i}^{2}$ and
$\zeta=\frac{27\alpha_{s}^{2}(Q^{2})}{160\mathcal{R}^{2}Q^{2}}$.
The correlation length $\mathcal{R}$  determines the size of the
non-linear terms. This value depends on how the gluon ladders are
coupled to the nucleon or on how the gluons are distributed within
the nucleon. The $\mathcal{R}$ is approximately equal to $\simeq
5~\mathrm{GeV}^{-1}$ if the gluons are populated across the proton
and it is equal to $\simeq 2~\mathrm{GeV}^{-1}$ if the gluons have
hotspot like structure. Here the higher dimensional gluon
distribution(i.e., higher twist) is assumed to be zero.\\
Recently in Ref.[11] the nonlinear modification of the evolution
of the gluon density from the parametrization  of $F_{2}$ in the
leading order of perturbation theory is considered. To investigate
the role of the non-linear corrections on the behaviour of the
longitudinal structure function in the low $Q^{2}$ region, we
consider the GLR-MQ equation for singlet structure function and
the AM equation [12] for longitudinal structure function at small
$x$. The longitudinal structure function in Ref.[12] is defined by
\begin{eqnarray}
F_{L}(x,Q^{2})&=&C_{L,ns+s}{\otimes}F_{2}(x,Q^{2})\nonumber\\
&&+<e^{2}>C_{L,g}{\otimes}G(x,Q^{2}),
\end{eqnarray}
where the non-singlet densities become negligibly small in
comparison with the singlet densities at small $x$. The symbol
$\otimes$ denotes convolution according to the usual prescription.
The perturbative expansion of the coefficient functions can be
written as $$
C_{L,i}(a_{s},x)=\sum_{n=1}a_{s}^{n}(Q^{2})c_{l,i}^{(n)}(x)
$$
where $a_{s}(Q^{2})=\frac{\alpha_{s}(Q^{2})}{4\pi}$ and the
coefficient functions $c_{l,i}^{(n)}(x)$, where $n$ is the order
in the running coupling, can be find in [13]. Defining the Laplace
transform  of the structure and distribution functions explicitly
from (1) and (2), one obtains the structure and distribution
functions in $s$-space as
\begin{eqnarray}
{\mathcal{L}}[\mathcal{\widehat{F}}_{L}(\nu,Q^{2});s]&=&f_{L}(s,Q^{2}),\nonumber\\
{\mathcal{L}}[\mathcal{\widehat{F}}_{2}(\nu,Q^{2});s]&=&f_{2}(s,Q^{2}),\nonumber\\
{\mathcal{L}}[{\widehat{G}}(\nu,Q^{2});s]&=&g(s,Q^{2})
\end{eqnarray}
where
\begin{eqnarray}
\mathcal{\widehat{F}}_{L}(\nu,Q^{2})&=&F_{L}(e^{-\nu},Q^{2}),\nonumber\\
\frac{\partial{\mathcal{\widehat{F}}_{2}(\nu,Q^{2})}}{\partial{\ln}Q^{2}}&=&
\frac{{\partial}F_{2}(e^{-\nu},Q^{2})}{\partial{\ln}Q^{2}},\nonumber\\
{\widehat{G}}(\nu,Q^{2})&=&G(e^{-\nu},Q^{2}).
\end{eqnarray}
and the coordinate transformation of $\nu$ is defined as
$\nu{\equiv}\ln(1/x)$. Now one could rewrite Eqs. (1) and (2) in
$s$-space as
\begin{eqnarray}
\frac{\partial{f_{2}(s,Q^{2})}}{\partial{\ln}Q^{2}}&{\simeq}&
\Phi_{f}(s)f_{2}(s,Q^{2})+<e^{2}>\Theta_{f}(s)g(s,Q^{2})\nonumber\\
&&-<e^{2}>\zeta g^{2}(s,Q^{2}),
\end{eqnarray}
and
\begin{eqnarray}
 f_{L}(s,Q^{2})=\Phi_{L}(s)f_{2}(s,Q^{2})
+<e^{2}>\Theta_{L}(s)g(s,Q^{2}).
\end{eqnarray}
In Eqs. (5) and (6) we used the fact that the Laplace transform of
a convolution function is simply  ordinary product of the Laplace
transform of that function. Transformation of Eq.(1) to $s$-space
in Eq.(5) is defined in a limited approach as in $\nu$-space we
assume that the Laplace transform of the
${\mathcal{L}}[\widehat{G}^{2}(\nu,Q^{2});s]$ to be less than
$g^{2}(s,Q^{2}){\equiv}[G(s,Q^{2})]^2$. Indeed
${\mathcal{L}}[\widehat{G}^{2}(\nu,Q^{2});s]<{\mathcal{L}}[\widehat{G}(\nu,Q^{2});s]^{2}
$.\footnote{The standard parametrization  of the gluon
distribution function at low $x$ introduced by $$
G(x,Q^{2})=f(Q^{2})x^{-\delta}
$$ where the low $x$ behavior could well be more singular. By considering the variable change
$\nu{\equiv}\ln(1/x)$, one can rewrite the gluon distribution in
$s$-space as
\begin{eqnarray}
{\mathcal{L}}[\widehat{G}^{2}(\nu,Q^{2});s]{=}\frac{f(Q^{2})^{2}}{(s-2\delta)},\nonumber\\
{\mathcal{L}}[\widehat{G}(\nu,Q^{2});s]^{2}{=}\frac{f(Q^{2})^{2}}{(s-\delta)^{2}}.\nonumber
\end{eqnarray}
We observe that the function
${\mathcal{L}}[\widehat{G}^{2}(\nu,Q^{2});s]$  is always lower
than ${\mathcal{L}}[\widehat{G}(\nu,Q^{2});s]^{2}$ for low $s$
values in a wide range of $Q^{2}$ values. According to this
result, we use from this limited approach for solving the
quadratic equation in $s$-space.} Therefore, the non-linear
corrections to the  longitudinal structure function is defined
into the proton structure function and the derivative of the
proton structure function with respect to $\ln{Q^{2}}$ in
$s$-space by a quadratic equation.  Inserting Eq. (5) in to Eq.
(6) the longitudinal structure function becomes
\begin{eqnarray}
f_{L}^{2}(s,Q^{2})-k(s,Q^{2})f_{L}(s,Q^{2})+h(s,Q^{2})=0,
\end{eqnarray}
where
\begin{eqnarray}
k(s,Q^{2})&=&\eta \Theta_{L}(s,Q^{2})\Theta_{f}(s,Q^{2})+2\Phi_{L}(s,Q^{2})f_{2}(s,Q^{2})\nonumber\\
h(s,Q^2)&=&\eta
\Theta_{L}(s,Q^{2})\Theta_{f}(s,Q^{2})\Phi_{L}(s,Q^{2})f_{2}(s,Q^{2})\nonumber\\
&&-\eta \Theta_{L}^{2}(s,Q^{2})
\Phi_{f}(s,Q^{2})f_{2}(s,Q^{2})\nonumber\\
 &&+\Phi_{L}^{2}(s,Q^{2})f_{2}^{2}(s,Q^{2})\nonumber\\
 &&+\eta
\Theta_{L}^{2}(s,Q^{2})Df_{2}(s,Q^{2}),\nonumber\\
\end{eqnarray}
where $$\eta=<e^{2}>/\zeta$$ and
$$Df_{2}(s,Q^{2})={\partial{f_{2}(s,Q^{2})}}/{\partial{\ln}Q^{2}}.$$
The coefficient functions $\Phi$ and $\Theta$ in $s$-space at
leading-order (LO) approximation are given by
\begin{eqnarray}
\Phi_{L}(s,Q^{2})&=&\frac{\alpha_{s}(Q^{2})}{\pi}C_{F}\frac{1}{2+s},\nonumber\\
\Theta_{L}(s,Q^{2})&=&2n_{f}\frac{\alpha_{s}(Q^{2})}{\pi}(\frac{1}{2+s}-\frac{1}{3+s}),\nonumber\\
\Theta_{f}(s,Q^{2})&=&n_{f}\frac{\alpha_{s}(Q^{2})}{2\pi}(\frac{1}{1+s}-\frac{2}{2+s}+\frac{2}{3+s}),\nonumber\\
\Phi_{f}(s,Q^{2})&=&\frac{\alpha_{s}(Q^{2})}{4\pi}[4-\frac{8}{3}(\frac{1}{1+s}+\frac{1}{2+s}\nonumber\\
&&+2(\psi(s+1)+\gamma_{E}))],
\end{eqnarray}
where $\psi(x)$ is the digamma function and $\gamma_{E}=0.5772156
. .$ is Euler constant. For the SU(N) gauge group, $C_{F}=4/3$
is the color Cassimir operator in QCD.\\
The quadratic equation (7) has two roots. Consequently the
longitudinal structure function in $s$-space is given by
\begin{eqnarray}
f_{L}(s,Q^{2})=\frac{1}{2}k(s,Q^{2})[1{\pm}(1-\frac{4h(s,Q^{2})}{k^{2}(s,Q^{2})})^{1/2}
].
\end{eqnarray}
The above equation (i.e., Eq.(10)) can be solving by a Taylor
series expansion around a particular choice of a point of
expansion, as the series is convergent when
$\beta=\frac{4h}{k^{2}}<1$.
\begin{figure}[h]
\includegraphics[width=0.45\textwidth]{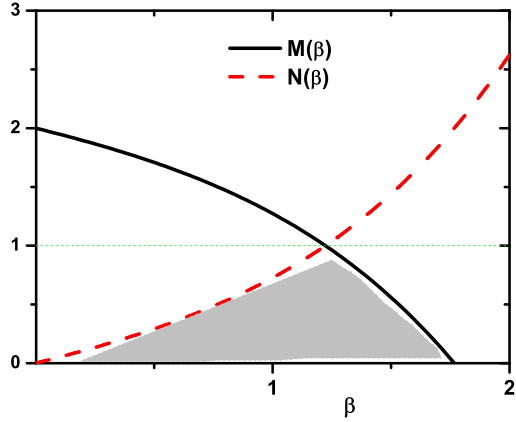}
\caption{The shaded area represents the convergence of positive
($M(\beta)$- solid curve) and negative ($N(\beta)$-dashed curve)
roots in Eq.(11).}\label{Fig1}
\end{figure}
Therefore, the non-linear longitudinal structure function  has the
following forms:
\begin{eqnarray}
f_{L}(s,Q^{2})=\frac{1}{2}k(s,Q^{2})\left\{ \begin{array}{rcl}
 M(\beta) & \mbox{for~positive~root} \\ N(\beta) & \mbox{for~negative~root}
 \end{array}\right.
\end{eqnarray}
where
$$M(\beta)=2-\frac{1}{2}\beta-\frac{1}{8}\beta^{2}-\frac{1}{16}\beta^{3}-\frac{5}{128}\beta^{4}..
$$
and
$$N(\beta)=\frac{1}{2}\beta+\frac{1}{8}\beta^{2}+\frac{1}{16}\beta^{3}+\frac{5}{128}\beta^{4}..
$$
In Fig.1 the behavior of $M(\beta)$ and $N(\beta)$ with respect to
the $\beta$ function are shown. In this figure the shaded area is
show the convergence region for the functions $M(\beta)$ and
$N(\beta)$. Thus, we rewrite Eq.(11) until the fourth sentence as
\begin{eqnarray}
f_{L}(s,Q^{2})=\left\{ \begin{array}{rcl}
 k-\frac{h}{k}-\frac{h^{2}}{k^{3}}-2\frac{h^{3}}{k^{5}}-5\frac{h^{4}}{k^{7}}  \\
  \frac{h}{k}+\frac{h^{2}}{k^{3}}+2\frac{h^{3}}{k^{5}}+5\frac{h^{4}}{k^{7}}
 \end{array}\right.
\end{eqnarray}
The calculation of the above sentences in Eq.(12) and the inverse
Laplace transforms of those are straightforward, and we find that
\begin{widetext}
\begin{eqnarray}
\widehat{J}_{0}(\nu){\equiv}{\mathcal{L}}^{-1}[k(s,Q^{2});\nu]&=&(\frac{16\alpha_{s}^2}{9\pi^2}\eta
+\frac{4\alpha_{s}}{3\pi}\widehat{F}_{2}(\nu,Q^{2}))\delta(\nu),\\
\widehat{J}_{1}(\nu){\equiv}{\mathcal{L}}^{-1}[\frac{h(s,Q^{2})}{k(s,Q^{2})};\nu]&=&
(\frac{16\alpha_{s}^2}{9\pi^2}\eta
\widehat{DF}_{2}(\nu,Q^{2})+\frac{64\alpha_{s}^3}{27\pi^3}\eta(-\ln(2)+\frac{1}{2})\widehat{F}_{2}(\nu,Q^{2})
+\frac{32\alpha_{s}^3}{27\pi^3}\eta\widehat{F}_{2}(\nu,Q^{2})
+\frac{4\alpha_{s}^2}{9\pi^2}\widehat{F}^{2}_{2}(\nu,Q^{2}))\nonumber\\
&&{\times}(\frac{16\alpha_{s}^2}{9\pi^2}\eta
+\frac{4\alpha_{s}}{3\pi}\widehat{F}_{2}(\nu,Q^{2}))^{-1}\delta(\nu),\nonumber\\
\widehat{J}_{2}(\nu){\equiv}{\mathcal{L}}^{-1}[\frac{h^{2}(s,Q^{2})}{k^{3}(s,Q^{2})};\nu]&=&
(\frac{16\alpha_{s}^2}{9\pi^2}\eta
\widehat{DF}_{2}(\nu,Q^{2})+\frac{64\alpha_{s}^3}{27\pi^3}\eta(-\ln(2)+\frac{1}{2})\widehat{F}_{2}(\nu,Q^{2})
+\frac{32\alpha_{s}^3}{27\pi^3}\eta\widehat{F}_{2}(\nu,Q^{2})
+\frac{4\alpha_{s}^2}{9\pi^2}\widehat{F}^{2}_{2}(\nu,Q^{2}))^{2}\nonumber\\
&&{\times}(\frac{16\alpha_{s}^2}{9\pi^2}\eta
+\frac{4\alpha_{s}}{3\pi}\widehat{F}_{2}(\nu,Q^{2}))^{-3}\delta(\nu),\nonumber\\
\widehat{J}_{3}(\nu){\equiv}{\mathcal{L}}^{-1}[\frac{2h^{3}(s,Q^{2})}{k^{5}(s,Q^{2})};\nu]&=&
(\frac{16\alpha_{s}^2}{9\pi^2}\eta
\widehat{DF}_{2}(\nu,Q^{2})+\frac{64\alpha_{s}^3}{27\pi^3}\eta(-\ln(2)+\frac{1}{2})\widehat{F}_{2}(\nu,Q^{2})
+\frac{32\alpha_{s}^3}{27\pi^3}\eta\widehat{F}_{2}(\nu,Q^{2})
+\frac{4\alpha_{s}^2}{9\pi^2}\widehat{F}^{2}_{2}(\nu,Q^{2}))^{3}\nonumber\\
&&{\times}(\frac{16\alpha_{s}^2}{9\pi^2}\eta
+\frac{4\alpha_{s}}{3\pi}\widehat{F}_{2}(\nu,Q^{2}))^{-5}\delta(\nu),\nonumber\\
\widehat{J}_{4}(\nu){\equiv}{\mathcal{L}}^{-1}[\frac{5h^{4}(s,Q^{2})}{k^{7}(s,Q^{2})};\nu]&=&
(\frac{16\alpha_{s}^2}{9\pi^2}\eta
\widehat{DF}_{2}(\nu,Q^{2})+\frac{64\alpha_{s}^3}{27\pi^3}\eta(-\ln(2)+\frac{1}{2})\widehat{F}_{2}(\nu,Q^{2})
+\frac{32\alpha_{s}^3}{27\pi^3}\eta\widehat{F}_{2}(\nu,Q^{2})
+\frac{4\alpha_{s}^2}{9\pi^2}\widehat{F}^{2}_{2}(\nu,Q^{2}))^{4}\nonumber\\
&&{\times}(\frac{16\alpha_{s}^2}{9\pi^2}\eta
+\frac{4\alpha_{s}}{3\pi}\widehat{F}_{2}(\nu,Q^{2}))^{-7}\delta(\nu).\nonumber
\end{eqnarray}
\end{widetext}
Therefore the non-linear corrections to the longitudinal structure
function due to the Laplace-transform method is defined by the
parametrization of $F_{2}(x,Q^{2})$ and its derivative
$DF_{2}(x,Q^{2})$ at low $x$ as we have
\begin{eqnarray}
F_{L}(x,Q^{2})&=& J_{0}(x,Q^{2})
+\sum_{n=0}^{\infty}J_{n+1}(x,Q^{2}),
\end{eqnarray}
where
\begin{eqnarray}
{J}_{0}(x,Q^{2})&=&(\frac{16\alpha_{s}^2}{9\pi^2}\eta
+\frac{4\alpha_{s}}{3\pi}{F}_{2}(x,Q^{2})),\\
{J}_{n+1}(x,Q^{2})&=&(\frac{16\alpha_{s}^2}{9\pi^2}\eta
{DF}_{2}(x,Q^{2})+\frac{32\alpha_{s}^3}{27\pi^3}\eta{F}_{2}(x,Q^{2})\nonumber\\
&&+\frac{64\alpha_{s}^3}{27\pi^3}\eta(-\ln(2)+\frac{1}{2}){F}_{2}(x,Q^{2})\nonumber\\
&&+\frac{4\alpha_{s}^2}{9\pi^2}{F}^{2}_{2}(x,Q^{2}))^{n+1}\nonumber\\
&&{\times}(\frac{16\alpha_{s}^2}{9\pi^2}\eta
+\frac{4\alpha_{s}}{3\pi}{F}_{2}(x,Q^{2}))^{-(2n+1)},\nonumber
\end{eqnarray}
where
\begin{eqnarray}
F_{ 2}(x,Q^{2})& =& D(Q^{2})(1-
x)^{n}\sum_{m=0}^{2}A_{m}(Q^{2})L^{m},
\end{eqnarray}
and
\begin{eqnarray}
DF_{2}(x,Q^{2}){\equiv}\frac{{\partial}F_{2}(x,Q^{2})}{\partial{\ln}Q^{2}}&=&
F_{2}(x,Q^{2})[\frac{{\partial}{\ln}D(Q^{2})}{\partial{\ln}Q^{2}}\nonumber\\
&&+\frac{{\partial}{\ln}\sum_{m=0}^{2}A_{m}(Q^{2})L^{m}}{\partial{\ln}Q^{2}}].\nonumber
\end{eqnarray}
The explicit expression for the proton structure function and
effective parameters are defined in Appendix A and Table I. Now,
with the explicit form of the proton structure function, we can
proceed to extract the non-linear corrections to the longitudinal
structure function $F_{L}(x,Q^{2})$  from data mediated by the
parametrization of $F_{ 2}(x,Q^{2})$ and its derivative. In Fig.2,
we show the $Q^{2}$-dependence of the non-linear corrections to
the longitudinal structure function at low $x$. Results of
calculations and comparison with the H1 collaboration data [14]
are presented in this figure (i.e., Fig.2), where the charm quark
mass effects are considered in the rescaling variable. These
results have been performed at fixed value of the invariant mass
$W$ as $W=230~ \mathrm{GeV}$.  The extracted non-linear
longitudinal structure functions are in good agreement in
comparison with the H1 collaboration data over a wide range of
 $Q^{2}$ values. Also this behavior improved
at low values of $Q^{2}$ in comparison with the other
parametrization models. We observe that at low $Q^{2}$, the
behavior of the non-linear longitudinal structure functions is
reduced in comparison with the behavior of the linear. In Fig.3 we
obtained our results with respect to the only negative roots and
compared with other results as defined in Fig.2. In these figures
(i.e., Fig.2 and Fig.3) we observe that the positive roots
decrease the growth rate of $F_{L}$ as $Q^{2}$ decreases. However,
the high-order corrections are important for comparing these
results with experimental
data at low $Q^{2}$. \\

In conclusion, we have presented a certain theoretical model to
describe the non-linear corrections to the longitudinal structure
function based on the Laplace transform method  at low values of
$x$ in a limited approach. A detailed analysis has been performed
to find an analytical solution of the non-linear longitudinal
structure function in $s$-space into the proton structure function
and its derivative. The nonlinear corrections improved the
behavior of the longitudinal structure function at low values of
$Q^{2}$ in comparison with the H1 collaboration data, but  the
high-order
corrections are important in the region of low $Q^{2}$.\\

\subsection{ACKNOWLEDGMENTS}

The author is thankful to the Razi University for financial
support of this project.\\

\subsection{Appendix A}
The proton structure function parameterized in Ref.[3] provide
good fits to the HERA data at low $x$ and large $Q^{2}$ values.
The explicit expression for the proton structure function, with
respect to the Block-Halzen fit, in a range of the kinematical
variables $x$ and $Q^{2}$, $x{\leq}0.1$ and
$0.15~\mathrm{GeV}^{2}<Q^{2}<3000~\mathrm{GeV}^{2}$, is defined by
the following form
\begin{eqnarray}
F^{\gamma p}_{ 2}(x,Q^{2})& =& D(Q^{2})(1-
x)^{n}[C(Q^{2})+A(Q^{2})\ln(\frac{1}{x}\frac{Q^{2}}{Q^{2}+\mu^{2}})\nonumber\\
&&+B(Q^{2})\ln^{2}(\frac{1}{x}\frac{Q^{2}}{Q^{2}+\mu^{2}})],
\end{eqnarray}
where
\begin{eqnarray}
 A(Q^{2})& =& a_{0} + a_{1} {\ln}(1+\frac{Q^{2}}{\mu^{2}}) + a_{2}{\ln}^{2}(1+\frac{Q^{2}}{\mu^{2}})
 ,\nonumber\\
B(Q^{2})& =& b_{0} + b_{1} {\ln}(1+\frac{Q^{2}}{\mu^{2}}) +
b_{2}{\ln}^{2}(1+\frac{Q^{2}}{\mu^{2}})
 ,\nonumber\\
C(Q^{2})& =& c_{0} + c_{1}
{\ln}(1+\frac{Q^{2}}{\mu^{2}}),\nonumber\\
D(Q^{2})& =& \frac{Q^{2}(Q^{2}+\lambda M^{2})}{(Q^{2}+M^{2})^2}.
\end{eqnarray}
Here $M$ is the effective mass and $\mu^{2}$ is a scale factor.
The additional parameters with their statistical errors are given
in Table I.\\
\begin{table}[h]
\caption{ The effective parameters at low $x$ for
$0.15~\mathrm{GeV}^{2}<Q^{2}<3000~\mathrm{GeV}^{2}$ provided by
the following values. The fixed  parameters are defined by the
Block-Halzen fit to the real photon-proton cross section as
$M^{2}=0.753 \pm 0.068~ \mathrm{GeV}^{2}$, $\mu^2 = 2.82 \pm
0.290~ \mathrm{GeV}^{2}$ and $c_{0} = 0.255 \pm 0.016$ [3].}
\begin{tabular} {cccc}
\toprule \\  \multicolumn{2}{c}{parameters \quad \quad \quad ~~~~~~~~~~~~~~~~value}    \\ &&&\\ \hline \\ &&&\\
  $a_{0} $  &   \quad  $8.205\times 10^{-4}~~  \pm  4.62\times10^{-4} $  \\

  $a_{1} $  &   \quad   $-5.148\times 10^{-2}\pm 8.19\times10^{-3}$  \\

  $a_{2}$   &    \quad  $-4.725\times 10^{-3}\pm 1.01\times10^{-3}$   \\  &&&\\

 $b_{0}$   &   \quad   $2.217\times 10^{-3}\pm 1.42\times10^{-4} $ \\

 $b_{1}$   &   \quad   $1.244\times 10^{-2}\pm 8.56\times10^{-4}$  \\

 $b_{2}$    &    \quad  $5.958\times 10^{-4}\pm 2.32\times10^{-4} $ \\ &&& \\

$c_{1}$& \quad  $1.475\times 10^{-1}~\pm 3.025\times10^{-2}$ & &\\

$n$& \quad  $11.49\pm 0.99$ & &\\

$\lambda$& \quad  $2.430~\pm 0.153$ & &\\

$\chi^{2}(\mathrm{goodness~ of~ fit})$ &  \quad  $0.95$ & &\\
\hline

\end{tabular}
\end{table}
\begin{figure}
\includegraphics[width=0.55\textwidth]{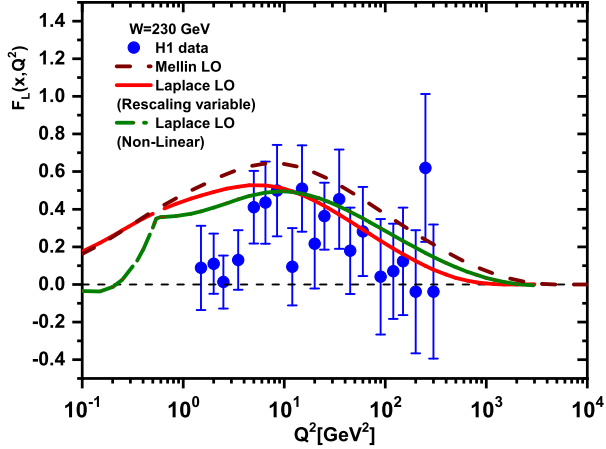}
\caption{Non-linear corrections to the $Q^{2}$ dependence of the
extracted longitudinal structure function  at fixed value of the
invariant mass $W=230~ \mathrm{GeV}$ (short-dashed curve) compared
with the Mellin transform method [1](dashed curve) at the LO
approximation and also the Laplace transform method [2](solid
curve) with the rescaling variable at the LO approximation.
Experimental data by the H1 Collaboration are taken from Ref. [14]
as accompanied with total errors.}\label{Fig2}
\end{figure}
\begin{figure}
\includegraphics[width=0.55\textwidth]{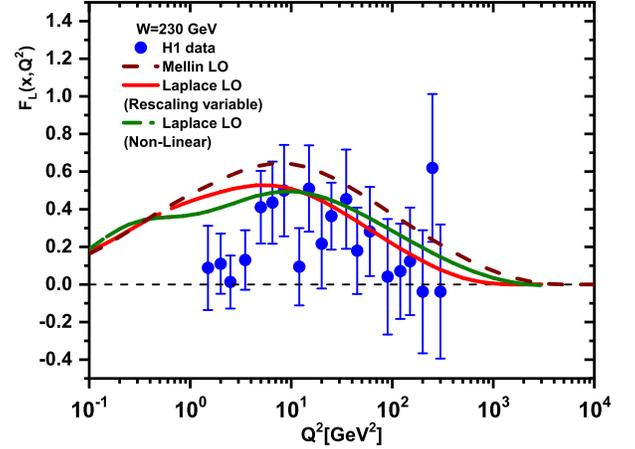}
\caption{Non-linear corrections to the $Q^{2}$ dependence of the
extracted longitudinal structure function, with respect to the
only negative roots, at fixed value of the invariant mass $W=230~
\mathrm{GeV}$ (short-dashed curve) compared with the Mellin
transform method [1](dashed curve) at the LO approximation and
also the Laplace transform method [2](solid curve) with the
rescaling variable at the LO approximation. Experimental data by
the H1 Collaboration are taken from Ref. [14] as accompanied with
total errors.}\label{Fig2}
\end{figure}
\section{References}

1. L.P.Kaptari et al., JETP Lett.{\bf 109}, 281(2019).\\
2. G.R.Boroun, arXiv:2108.09465.\\
3. M. M. Block, L. Durand and P. Ha, Phys. Rev. D {\bf89}, 094027
(2014).\\
4. F. D. Aaron et al. (H1 and ZEUS Collaborations), JHEP
{\bf1001}, 109 (2010).\\
5. H1 and ZEUS Collaborations (Abramowicz H. et al.), Eur. Phys.
J. C {\bf78}, 473 (2018).\\
6.M.R.Pelicer et al., Eur.Phys.J.C {\bf79}, 9 (2019).\\
7. B.Rezaei and G.R.Boroun, Phys.Lett.B {\bf692}, 247 (2010); Phys.Rev.C {\bf101}, 045202 (2020).\\
8. M.Devee and J.K.Sarma, Eur.Phys.J.C {\bf74}, 2751 (2014);
M.Devee, arXiv [hep-ph]: 1808.00899 (2018); P.Phukan, M.Lalung and
J.K.Sarma, Nucl.Phys.A {\bf968}, 275 (2017).\\
9. G.R.Boroun, Eur.Phys.J.A {\bf43}, 335 (2010); G.R.Boroun and B.Rezaei, arXiv:2107.11033.\\
10. L.V.Gribov, E.M.Levin and M.G.Ryskin, Phys.Rept.{\bf100}, 1
(1983); A.H.Mueller and J.w.Qiu, Nucl.Phys.B {\bf268},
427 (1986).\\
11.A.V.Kotikov, JETP Lett.{\bf111}, 67 (2020).\\
12. G.Altarelli and G.Martinelli, Phys.Lett.B\textbf{76}, 89(1978).\\
13. S. Moch, J.A.M. Vermaseren, and A. Vogt, Phys. Lett. B
{\bf606}, 123 (2005).\\
14. H1 Collab. (V.Andreev , A.Baghdasaryan, S.Baghdasaryan et
al.), Eur.Phys.J.C{\bf74},
2814(2014).\\


\end{document}